# Fuzzy Petri Nets for Human Behavior Verification and Validation

M. Kouzehgar, M. A. Badamchizadeh, S. Khanmohammadi

Department of Control Engineering
Faculty of Electrical and Computer Engineering, University of Tabriz,
Tabriz, Iran

*Abstract*— **Regarding the rapid growth of the size and complexity of simulation applications, designing applicable and affordable verification and validation (V&V) structures is an important problem. On the other hand, nowadays human behavior models are principles to make decision in many simulations and in order to have valid decisions based on a reliable human decision model, first the model must pass the validation and verification criteria. Usually human behavior models are represented as fuzzy rule bases. In all the recent works, V&V process is applied on a ready given rule-base. In this work, we are first supposed to construct a fuzzy rule-base and then apply the V&V process on it. Considering the professor-student interaction as the case-study, in order to construct the rule base, a questionnaire is designed in a special way to be transformed to a hierarchical fuzzy rule-base. The constructed fuzzy rule base is then mapped to a fuzzy Petri net and then within the verification (generating and searching the reachability graph) and validation (reasoning the Petri net) process is searched for probable structural and semantic errors.**

## I. INTRODUCTION

All On the whole nowadays human behavior models are principles to make decision in many simulations. In order to improve the fidelity and automation of simulation exercises, human behavior models have become key components in most simulations. Our work firstly suggests the setup of the students' deduction system and looking at the student-professor interaction as a system; furthermore the application of verification and validation in education system is tested. The professor-student interaction is chosen as the case-study since it is something tangible for all of us who really live in an academic environment being in direct contact with such a system in our everyday life and something you must deal with a day after the other really was worth giving a try.

It's high time simulating human behavior and presenting relevant control schemes has become an exciting field for the researchers. Human behavior modeling or human behavior representation (HBR) is a field of study important in military service research [1, 2], robotics [3], brain-computer interface (BCI) , human machine interface (HMI) [4, 5] and some specially oriented anthropology studies [6].Human behavior models are often represented by finite state machines, rules, fuzzy rules [7], artificial neural networks [8], fuzzy hybrid rule-frames [9], fuzzy dynamic Bayesian networks [10, 11], concept lattice [12], multi-agent based modeling [13]. Among all these, HBR by a fuzzy rule base is the most common [14].

In order to have a reliable human behavior model on which many decisions depends, it is essential to ensure that the HBM passes through V&V criteria. In this research V&V for a fuzzy rule-base is focused. Some techniques for verification of rule based systems are presented in recent works. In [15] and [16], the rules are grouped into sets according to some criteria, and each rule within a set is statically compared to every other one to check consistency and completeness properties. In [17], within an exhaustive computationally expensive approach any chaining of rules is taken into account from which an inconsistency could be deduced. Despite [17], there are incremental approaches that check the rule base after each modification during development [18]. Some references use some graphical notations such as Petri nets to represent rules and detect the structural errors of rule bases. In [19], based on the concept of $\omega$-nets [20], a special reachability graph is presented to detect structural errors in rule bases. This technique is applied in later researches [1, 14]. In [21], a fuzzy rule base systems verification method using high-level Petri nets is discussed. Furthermore, in [1], a double-phase verification technique for HBM is presented that consists of weak and strong verification whereby [14] adds a solution to semantically validate the HBM. In summary, the existing V&V techniques for rule bases mainly focus on structural verification and rarely deal with validation issues, let alone semantic validation of special cases such as human behavior models. Furthermore, in [22, 23], general V&V problems of human behavior models and its possible techniques are illustrated.

As a rather new field in human behavior, the educational system is regarded as the case-study, with which we deal in everyday life. In pursue to some recent works on student's performance evaluation [24, 25], the idea of entering the Professor into the system was initiated. In order to set up a rule base for the fuzzy system of human behavior, the effective parameters on decision making in that special field must be initially identified. On the way to this goal during careful consults with anthropology experts for a long time, many procedures were suggested to identify these parameters, among which lies survey research, -a subset of which is the Delfi technique- action research and correlation research. Finally according to the experts' recommendation, based on the techniques of questionnaire designation [26, 27 and 28], a questionnaire was designed in a special way to be transformed to a fuzzy rule-base with uncertain fuzzy rules dealing with certainty factors. Then the specially designed questionnaire was handed among many students several times and after each time the necessary changes was made on it in order to mostly satisfy the students' (SME's) points of view. Then the constructed fuzzy rule base is mapped to a fuzzy Petri net and afterwards the corresponding special reachability graph is generated and searched in order to distinguish errors dealing with verification. Then by means of a rule referent gathered from the subject matter experts' (SME) point of view-here the student's point of view- the rule base is semantically validated.

The paper is organized as follows: section II deals with the errors concerning human behavior models. Section III concerns itself with the introduction of fuzzy Petri nets and mapping the rule base to fuzzy Petri net. Section IV is dedicated to the introduction of the case study. Section V and VI respectively illustrate the V&V processes. Section VII concludes the paper and presents ideas for future works.

## II. FUZZY PETRI NETS

### A. Fuzzy Petri Nets- a brief Introduction

A fuzzy Petri net model (FPN) can be used to represent a fuzzy rule-based system. A FPN [19, 29] is a directed graph containing two types of nodes: places and transitions, where circles represent places and bars represent transitions. Each place represents an antecedent or consequent and may or may not contain a token associated with a truth degree between zero and one which speaks for the amount of trust in the validity of the antecedent or consequent. Each transition representing a rule is associated with a certainty factor value between zero and one. The certainty factor represents the strength of the belief in the rule. The relationships from places to transitions and vice versa are represented by directed arcs. The concept of FPN is derived from Petri nets. As with [14 and its refs], a generalized FPN structure can be defined as an 8-tuple:

FPN = (P, T, D, I, O, μ, α, β), where

P = {$p_1$, $p_2$, …,$p_n$} is a finite set of places,

T = {$t_1$, $t_2$, …,$t_m$} is a finite set of transitions,

D = {$d_1$, $d_2$, …,$d_n$} is a finite set of propositions ,

P ∩ T ∩ D = ∅, |P| = |D|

I: P × T → {0, 1} is the input function, a mapping from places to transitions;

O: T × P → {0, 1} is the output function, a mapping from places to transitions,

μ : T → [0,1] is an association function, a mapping from transitions to [0,1] i.e. the certainty factor α: P → [0,1] is an association function, a mapping from places to [0,1] i.e. the truth degree β: P → D , is an association function, a mapping from places to propositions.

### B. Mapping The Rule Base to FPN

During this mapping procedure, each rule is represented as a transition with its corresponding certainty factor and each antecedent is modeled by an input place and the consequents are modeled by out places with corresponding truth degrees. In this modeling a transition- here a rule- is enabled to be fired if all its input places have a truth degree equal to or more than a predefined threshold value [30]. As illustrated in Fig.1, after firing the rule, the output places will have a truth degree equal to the input place truth degree multiplied by the transition certainty factor [30].

In order to transform compound rules to FPNs, we first apply normalization rules introduced in [21] to change any rules to Horn Clauses [21 and its refs]. A Horn Clause is a kind of rule in the following form.

$$P_1 \wedge P_2 \wedge P_3 \wedge \dots P_{j-1} \to P_j$$

### III. ERRORS CONCERNING HUMAN BEHAVIOR MODELS

Human behavior models may suffer from two types of errors [14]. If modeled as a rule-base, they may suffer from the structural errors from which any fuzzy rule-base may suffer, among which incompleteness, inconsistency, circularity and redundancy are the most popular. On the other hand, HBMs as models being in contact with human operators and users, must meet the user's need. Any contrast with the user's point of view will be indicating the semantic errors dividing into two groups: semantic incompleteness and semantic incorrectness.

### A. Structural errors

As illustrated in [14, 21] the structural errors from which a rule base may suffer are as follows.

#### 1) Incompleteness

Incompleteness rules result from missing rules in a rule base. An example of incompleteness rules is as follows.

$r_1$ :  → $P_1$
$r_2$: $P_1 \wedge P_3$ → $P_2$
$r_3$: $P_1$ → $P_4$
$r_4$: $P_2$ → .

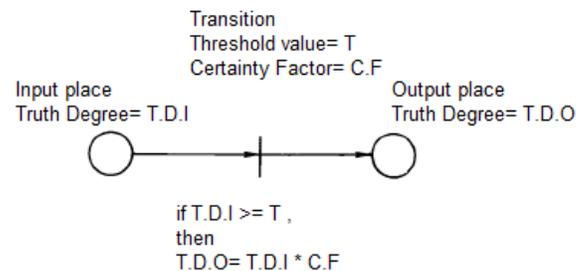

Figure 1.   Firing process

Rule $r_1$ is representing fact -a source transition in the FPN, while rule $r_4$ is representing a query- a sink transition in the FPN. Rule $r_2$ is a useless rule because the antecedent $P_3$ does not have a matching part appearing in the consequents of the rest of the rules, thus $P_3$ is a dangling antecedent. Rule $r_3$ is a useless rule because the consequent $P_4$ of $r_3$ does not have a matching part appeared in the antecedents of the rest of the rules, thus $p_4$ is a dead-end consequent.

### 2) Inconsistency

Inconsistency rules end in conflict and should be removed from the rule base. This means a set of rules are conflicting if contradictory conclusions can be derived under a certain condition. An example of inconsistency rules is as follows.

$r_1 : P_1 \wedge P_2 \rightarrow P_3$
$r_2 : P_3 \wedge P_4 \rightarrow P_5$
$r_3 : P_1 \wedge P_2 \wedge P_4 \rightarrow \sim P_5$

Rule $r_3$ is an inconsistent one because $P_1$ and $P_2$ ends in $P_3$ , $P_3$ and $P_4$ ends in $P_5$, while in $r_3$, $P_1$ and $P_2$ (the same $P_3$) and $P_4$ ends in ~$P_5$.

### 3) Circularity

Circular rules refer to the case that several rules have circular dependency. Circularity may end in an infinite reasoning loop and must be broken. An example of circular rules is as follows.

$r_1 : P_1 \rightarrow P_2$
$r_2 : P_2 \rightarrow P_3$
$r_3 : P_3 \rightarrow P_1$

### 4) Redundancy

Redundancy rules are unnecessary rules in a rule base. Redundancy rules increase the size of the rule based and may cause extra useless deductions. An example of redundancy rules is as follows.

$r_1 : P_1 \wedge P_3 \rightarrow P_2$
$r_2 : P_1 \wedge P_3 \rightarrow P_2$
$r_3 : P_1 \rightarrow P_2$
$r_4 : P_4 \rightarrow P_5$
$r_5 : P_4 \rightarrow P_5 \wedge P_6$

$r_1$ is the redundant rule of $r_2$. There are two cases of the directly subsumed rules. First, rules $r_1$ or $r_2$ is a subsumed rule of $r_3$ because $r_1$ or $r_2$ has more restrictive condition than $r_3$. Second, rule $r_4$ is a subsumed rule of $r_5$ because $r_4$ has less implied conclusion than $r_5$.

## B. . Semantic errors

As explained in [14] the semantic errors are classified into two levels.

### 1) Semantic incompleteness

In a human behavior model, semantic incompleteness happens if human behavior model does not meet users' requirements, and are reflected as missing rules, and missing antecedents or consequents in a rule from the users' point of view.

### 2) Semantic incorrectness

Semantic incorrectness occurs if the human decision model produces an output that is different from the expected output for given identical input data in the validation referent. Semantic incorrectness also indicates that the human behavior model doesn't meet the users' needs.

## IV. MODELING OF THE CASE- STUDY

In recent works [1, 14], V&V process is applied on a ready given rule-base. In this work we are first supposed to construct a rule-base and then apply the V&V process on it.

In order to make up a rule base for a human related case-study. The needed information is gathered through some filled in questionnaires whose questions specially designed in two stages that leads to a hierarchical fuzzy inference within two steps. of course it seems necessary to notify that the students who were supposed to fill in the questionnaires were not informed of the mentioned structure lying beyond.

The questions that form the first step of the hierarchical deduction are in fact the input properties of the fuzzy HBR system whose outcome serves the internal properties of the system. Finally in the second stage, the fuzzy deduction on the internal properties and some directly effective input properties makes up the second stage rule.

The questions standing for the input variables supposed to make a unique internal property are arranged in a diversified form in order not to impose any conditional effect on the participant's mind while answering the questions and to let the students answer the questions feeling absolutely free and keeping the system as fuzzy as possible.

The answers to the questions is given within a 5-optional list which includes selecting a linguistic variable (very low, low , medium, high, very high) of the fuzzy system. Also to get the truth degree of the antecedents and consequents and also the certainty factor for each rule, the participants are asked to answer the following question in terms of percent: *How much you are confident to your answer?*

Our present case study is defined on the decision factors a student considers while selecting a professor among others for a special presented course by several professors.

## A. The Input Properties

The system's input properties are gathered within 11 questions in the questionnaire. Questions 1 to 11 will be represented as Q1 to Q11 in the rest of paper. These questions are summarized as follows.

- Question 1-The professor's authority on the topic and his power to answer the questions?

- Question 2-Using references updated each semester and introducing further references to study?

- Question 3-How much the professor acts strong on presenting the topic?

- Question 4-How much he feels responsible for attending on time?

- Question 5-How much he optimally manages the time he has?

- Question 6- How much he appreciates the suggestions and constructive critics?

- Question 7-How much well-behaved he is?

- Question 8- Feeding the student with several quizzes and midterm exams?

- Question 9- presenting projects and assigning homework?

- Question 10- The range of marks?

- Question 11- Advice and suggestion by the elder students?

### B. The Internal Properties

The internal properties of the system are made on the basis of some composition of the input properties.

- The input properties Q1, Q2 and Q3 form an internal property called "The power of teaching".

- The input properties Q4 and Q5 form an internal property called "regularity".

- The input properties Q6 and Q7 form an internal property called "behavior".

- The input properties Q8 and Q9 form an internal property called "The power of attracting the student".

- The power of teaching, regularity, behavior, the power of attracting the student, elders' advice and the mark range are the factors making up the Prof's rank to be selected among others.

In other words, we have a fuzzy deduction in two levels: Level one is supposed to deduce the internal properties, level two is supposed to deduce the prof's rank based on the internal and input properties.

### 1) Level 1:

If Q1 is … and Q2 is … and Q3 is …, then the power of teaching is …

If Q4 is … and Q5 is …, then regularity is …

If Q6 is … and Q7 is … , then behavior is ....

If Q8 is … and Q9 is … , then the power of attracting the student is ....

### 2) Level 2:

If the power of teaching is …and regularity is … and behavior is …. and attractiveness is … then the prof's popularity is … .

Each of the blanks is filled with a linguistic value: very low, low, medium, high and very high. A sample rule base for the above case-study can be presented as a human behavior model (HBM) as the following structure shown in Fig. 2.

- HBM= (Prof-Student, IPS, InPS, OPS, RS);
-- HBM.IPS={Q1, Q2, Q3, Q4, Q5, Q6, Q7, Q8, Q9, Q10, Q11};
-- HBM.InPS={Tea, Reg, Beh, Att};
-- HBM.OPS={Pop};
-- HBM. RS={R1, R2, …, R10}

--- HBM.RS.R1=(Rule1, Q1(vh) ∧ Q2(h) ∧ Q3(vh), Tea(vh), 0.95);

--- HBM.RS.R2=(Rule2, Q4(h) ∧ Q5(h), Reg(h), 0.65);

--- HBM.RS.R3=(Rule3, Q6(m) ∧ Q7(h), Beh(h), 0.85);

--- HBM.RS.R4=(Rule4, Q8(m) ∧Q9(h), Att(m), 0.65);

--- HBM.RS.R5=(Rule5, Tea(vh) ∧ Reg(h) ∧Beh(h) ∧ Att(m) ∧ Mark(m) ∧Adv(h), Pop(h), 0.95);

--- HBM.RS.R6=(Rule6, Q1(m) ∧ Q2(vh) ∧ Q3(h), Tea(m), 0.80);

--- HBM.RS.R7=(Rule7, Q4(m) ∧Q5(h), Reg(vh), 0.6);

--- HBM.RS.R8=(Rule8, Q6(l) ∧Q7(h), Beh(vh), 0.85);

--- HBM.RS.R9=(Rule9, Q8(m) ∧Q9(h), Att(l), 0.75);

--- HBM.RS.R10=(Rule10, Tea(m) ∧ Reg(vh) ∧ Beh(vh) ∧ Att(l) ∧Mark(vh) ∧Adv(m), Pop( vh), 0.7);

In the above structure, human behavior model (HBM) is introduced into a 5-tuple consisting of the input property set (IPS), internal property set (InPS), output property set (OPS) and rule set (RS). Q1 to Q11 speak for Question 1 to Question 11 as input properties. Tea, Reg, Beh , Att and Pop respectively stand for the power of teaching, regularity, behavior, attractiveness and popularity as internal properties. Terms $vl$, $l$, $m$, $h$ and $vh$ respectively represent the linguistic values: very low, low, medium, high and very high. In the rules, the $2^{nd}$ element shows the antecedents, the $3^{rd}$ element shows the consequent and the last number shows the certainty factor dedicated to the rule. For example Rule1 is as follows.

- --- HBM.RS.R1=(Rule1, Q1(vh) ∧ Q2(h) ∧ Q3(vh), Tea(vh), 0.95);

- If Q1 is very high and Q2 is high and Q3 is very high, then the power of teaching is very high.

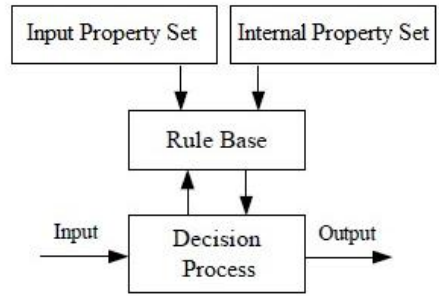

Figure 2.    The decision model

The corresponding Petri net model is illustrated in Fig. 3. In this Petri net model, according to the propositions dedicated to each place, transitions 1 to 10 respectively represent rules 1 to 10 in the introduced rule base above and firing each transition means the corresponding rule is fulfilled.

## V. VERIFICATION PROCESS

In order to fulfill the rule-base verification phase, we must first map the rule-base to Petri net as shown in Fig. 3. Then as with the algorithm mentioned in [19, 21] a special reachability graph is generated on the basis of the concept of ω-nets.

In this reachability graph, first, a zero vector is defined as the root node as long as the number of places. Then at any current marking, among the transitions yet not considered, the enabled transitions are determined. At each step by firing the set of enabled transitions, a new node is added to the graph in which the corresponding elements of the node- the places which are filled after firing the transitions- are set to ω which is assumed as a huge value. In this way at each step there's a marking. If firing of the transitions at a step ends in a repetitive marking, the graph will have a loop.

After generating the reachability graph, the structural errors including incompleteness, inconsistency, redundancy and circularity are distinguished. Then on the basis of the

SME point of view, the rule-base is verified to eliminate the errors.

The corresponding reachability graph for the above Petri net model is depicted in Fig. 4. The places P0 to P17 are regarded as TRUE antecedents and are initially filled (set to ω) for this reason. That's why in the first node there are 18 ω's. In this marking transitions T1, T2, T3, T4, T6, T7, T8 and T9 are enabled. After firing these transitions, in the second step, places P18 to P25 are filled and the corresponding values in the node vector are set to ω. On the final step by firing T5 and T10 (the enabled transitions), the places P26 and P27 will also be filled up.

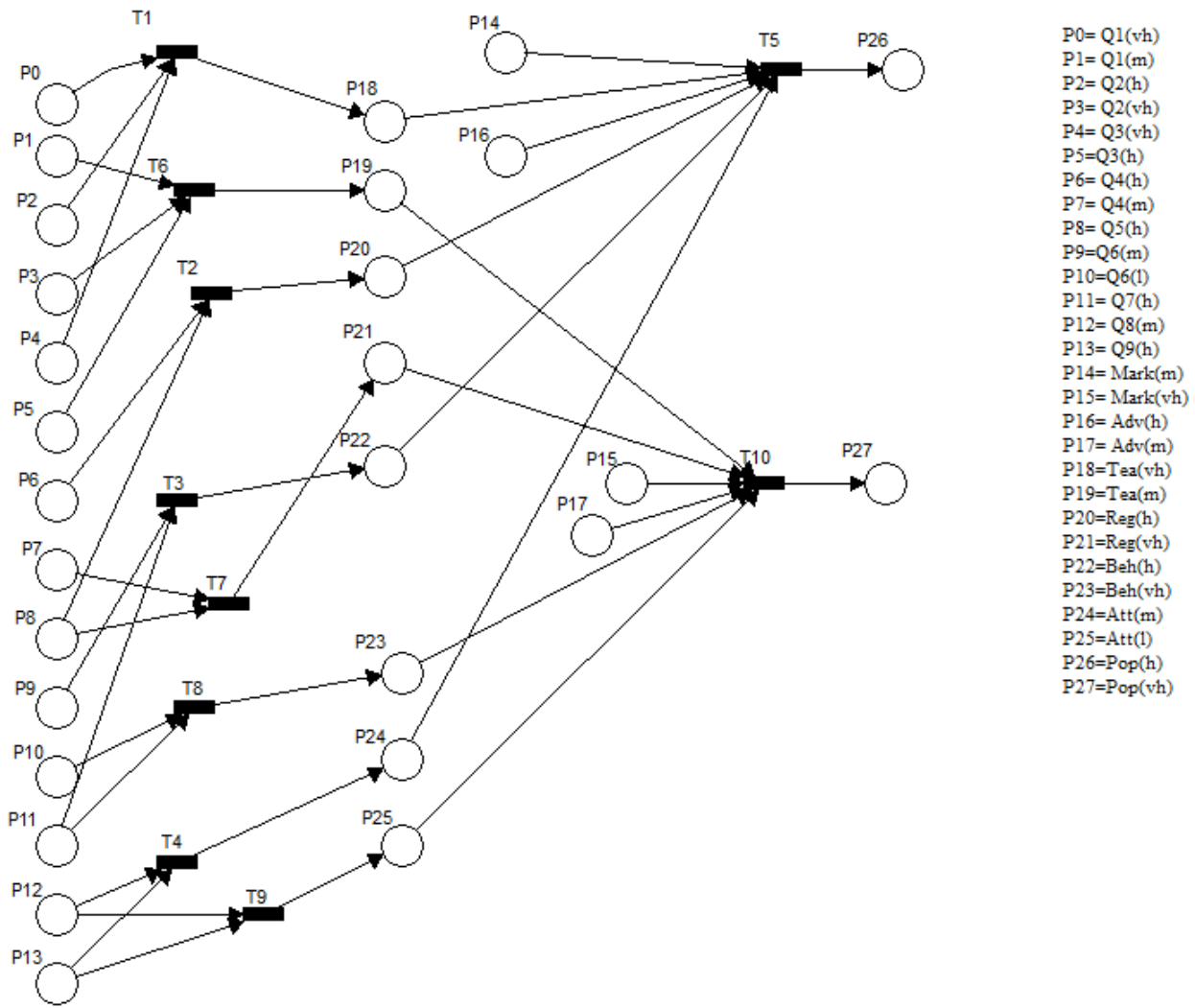

Figure 3. The Petri net representation of the HBM system

P0= Q1(vh)
P1= Q1(m)
P2= Q2(h)
P3= Q2(vh)
P4= Q3(vh)
P5=Q3(h)
P6= Q4(h)
P7= Q4(m)
P8= Q5(h)
P9=Q6(m)
P10=Q6(l)
P11= Q7(h)
P12= Q8(m)
P13= Q9(h)
P14= Mark(m)
P15= Mark(vh)
P16= Adv(h)
P17= Adv(m)
P18=Tea(vh)
P19=Tea(m)
P20=Reg(h)
P21=Reg(vh)
P22=Beh(h)
P23=Beh(vh)
P24=Att(m)
P25=Att(l)
P26=Pop(h)
P27=Pop(vh)

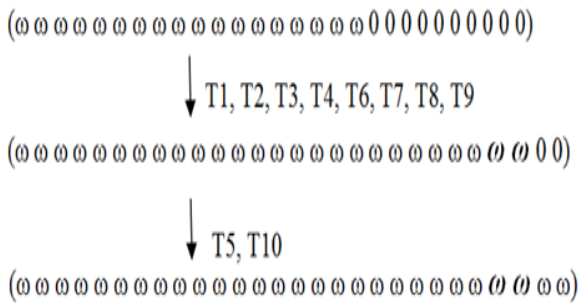

Figure 4. The reachability diagram

In the reachability graph shown in Fig. 4, all the places and transitions exist so there are no incompleteness errors. P24 and P25 are different states of one property (attractiveness). Their simultaneous existence may refer to some concept of inconsistency in the rule base. The reachability graph lacking any loops speaks for lacking circularity errors and finally having no transitions underlined, speaks for non-redundancy.

In order to verify the rule base after verification, the SME point of view must be considered to decide whether to omit or change a rule, or which rule must be changed or omitted to leave the rule base as a structurally fault-free one.

After a careful consult with the SME- here the students, we conclude that regarding the SME opinion, by omitting the Rule 9, the rule-base is refined after verification.

## VI. VALIDATION PROCESS

In order to fulfill the validation process, first of all a validation referent is required. In this research the information needed to construct a validation referent is gathered from the students' point of view - the SME of the present system.

Validation is carried out within two separate phases: static validation and dynamic validation [14].

### A. Static Validation

Static validation involves distinguishing the semantic incompleteness. In the static validation running or reasoning the FPN is not needed. In this phase only the places are searched and their properties are recorded and compared with the referent. If the number of searched input properties is less than the expected ones, the rule base may miss antecedents, if the number of searched output properties is less than the expected ones, the rule base may miss consequents and if the number of searched internal properties is less than the expected ones, the rule base may miss antecedents or consequents.

The referent for validation given by a student is as follows and the static validation results are summarized in Table1.

- HBM= (Prof-Student, IPSref, InPSref, OPSref, RSref);
-- HBM.IPSref={Q1, Q2, Q3, Q4, Q5, Q6, Q7, Q8, Q9, mark, advice};
-- HBM.InPSref={Tea, Reg, Beh, Att, scientific reputation};

-- HBM.OPSref={Pop};
-- HBM. RSref={Rref1, Rref2, …, Rref10}
--- HBM.RS.R1ref =(Rule1ref, Q1(vh) ∧ Q2(h) ∧ Q3(vh), Tea(vh), 0.95);
--- HBM.RS.R2ref =(Rule2ref, Q4(h) ∧ Q5(h), Reg(h), 0.65);
--- HBM.RS.R3ref =(Rule3ref, Q6(m) ∧ Q7(h), Beh(vh), 0.7);
--- HBM.RS.R4ref =(Rule4ref, Q8(m) ∧ Q9(h), Att(m), 0.65);
--- HBM.RS.R5ref =(Rule5ref, Tea(vh) ∧ Reg(h) ∧ Beh(h) ∧ Att(m) ∧ Mark(m) ∧ Adv(h), Pop( h), 0.95);
--- HBM.RS.R6ref =(Rule6ref, Q1(m) ∧ Q2(vh) ∧ Q3(h), Tea(m), 0.80);
--- HBM.RS.R7ref =(Rule7ref, Q4(m) ∧ Q5(h), Reg(h), 0.5);
--- HBM.RS.R8ref =(Rule8ref, Q6(l) ∧ Q7(h), Beh(vh), 0.85);
--- HBM.RS.R10ref =(Rule10ref, Tea(m) ∧ Reg(vh) ∧ Beh(vh) ∧ Att(l) ∧ Mark(vh) ∧ Adv(m), Pop(vh), 0.7);

According to Table1, the property of scientific reputation and rules R3ref and R7ref do not exist in the present rule-base. So the rule-base suffers from semantic incompleteness in this student's point of view. After interviewing many students, we conclude that the "scientific reputation" property can be neglected and the R7 is replaced with R7ref and R3 remains the same.

### B. Dynamic Validation

Dynamic validation involves clarifying the existence of semantic incorrectness through running and reasoning FPN. In order to fulfill the dynamic validation, the results of reasoning FPN for given inputs are compared to their counterparts in the validation referent to check if there's any semantic incorrectness.

As with [30], rules with certainty factors are classified into three types, among which we use the first type according to the nature of the existing rules.

- Type 1: if $P_1$ and $P_2$ and … $P_n$, then $P_m$.
- Type 2: if $P_n$ then $P_1$ and $P_2$… and $P_n$.
- Type 3: if $P_1$ or $P_2$… or $P_n$, then $P_m$.

If $\alpha_i$ is considered as the truth degree of antecedents or consequents and $\mu_i$ is the certainty factor dedicated to rule $r_i$, the rule and its uncertainty reasoning is as follows.

$$R_i: P_1(\alpha_1) \wedge P_2(\alpha_2) \wedge … P_{n-1}(\alpha_{n-1}) \rightarrow P_n(\alpha_n) \qquad CF = \mu_i$$
$$\alpha_n = \min\{\alpha_1, \alpha_2, …, \alpha_{n-1}\} \times \mu_i$$

In this part, the revised rule-base after the static validation must be considered.

TABLE I.    THE STATIC VALIDATION RESULTS

| Present FPN | | | | | Referent FPN | | | |
|---|---|---|---|---|---|---|---|---|
| IPS | InPS | OPS | RS-verified | | IPSref | InPSref | OPSref | RSref |
| Q1 | Teach | Pop | R1 | | Q1 | Teach | Pop | R1ref |
| Q2 | Regulation | | R2 | | Q2 | Regulation | | R2ref |
| Q3 | Behavior | | × | | Q3 | Behavior | | R3ref |
| Q4 | Attractiveness | | R4 | | Q4 | Attractiveness | | R4ref |
| Q5 | × | | R5 | | Q5 | Scientific Reputation | | R5ref |
| Q6 | | | R6 | | Q6 | | | R6ref |
| Q7 | | | × | | Q7 | | | R7ref |
| Q8 | | | R8 | | Q8 | | | R8ref |
| Q9 | | | R10 | | Q9 | | | R10ref |
| Mark | | | | | Mark | | | |
| Advice | | | | | Advice | | | |

The reference values for dynamic validation given by the student are as follows. The following numbers for reference values are gathered throughout the questionnaires by adding a choice to be filled in, in percent form.

-- reference values:

Ref − value 1: $\alpha(Q1(vh)) = 0.65 \ \wedge \ \alpha(Q2(h)) = 0.75 \ \wedge \ \alpha(Q3(vh)) = 0.9 \ \rightarrow \alpha(Tea(vh)) > 0.7$

Ref − value 2: $\alpha(Q4(h)) = 0.7 \ \wedge \alpha(Q5(h)) = 0.8 \ \rightarrow \alpha(Reg(h)) > 0.4$

Ref − value 3: $\alpha(Q1(vh)) = 0.65 \ \wedge \alpha(Q2(h))$
$= 0.75 \ \wedge \ \alpha(Q3(vh)) = 0.9 \ \wedge \alpha(Q4(h))$
$= 0.7 \ \wedge \ \alpha(Q5(h)) = 0.8 \ \wedge \ \alpha(Q6(m))$
$= 0.6 \ \wedge \ \alpha(Q7(h)) = 0.45 \ \wedge \alpha(Q8(m))$
$= 0.75 \ \wedge \alpha(Q9(h))$
$= 0.9 \ \wedge \ \alpha(mark(m))$
$= 0.95 \ \wedge \ \alpha(adv(h)) = 0.9 \ \rightarrow \ \alpha(Pop(h)) > 0.3$

Considering the certainty factors in the validation referent given above, by the reference values for reasoning we have:

Ref-value 1 is validated by the use of Rule1ref according to the correspondence between their antecedents and consequents. Minimum of the truth degrees of the antecedents given in the referent i.e. min(0.65,0.75,0.9), must be multiplied to the certainty factor given in the referent rule base (0.95) to obtain the truth degree of the consequent and compare it with the condition provided by the referent. As illustrated underneath, for this case, the validation criterion fails.

$min(0.65, 0.75, 0.9) \times 0.95 = 0.6175 < 0.7$
$\rightarrow semantic\ incorrectness, validation\ failed$

Similarly Ref-value 2 is validated by Rule2ref. Minimum of the truth degrees of the antecedents given in the referent i.e.

min (0.7, 0.8), is multiplied to the certainty factor given in the referent rule base (0.65) to obtain the truth degree of the consequent and is compared to the condition provided by the referent. For this case, the validation criterion is passed as follows.

$min(0.7, 0.8) \times 0.65 = 0.455 > 0.4$
$\rightarrow o.k! passed\ validation.$

In some cases the referent values maybe given in such a way that it is needed to merge rules during validation. In order to validate the Ref-value 3, according to the truth degrees given for special antecedents, rules Rule1ref to Rule5ref must be merged during validation first to obtain the truth degrees for the antecedents of Rule5ref.

- validation through Rule1ref to obtain the truth degree for Tea(vh) gives: $min(0.65, 0.75, 0.9) \times 0.95 = 0.6175$

- validation through Rule2ref to obtain the truth degree for Reg(h) gives: $min(0.7, 0.8) \times 0.65 = 0.455$

- validation through Rule3ref to obtain the truth degree for Beh(vh) gives: $min(0.6, 0.45) \times 0.7 = 0.315$

- validation through Rule4ref to obtain the truth degree for Att(m)gives: $min(0.75, 0.9) \times 0.65 = 0.4875$

- and finally validation through Rule5ref to obtain the truth degree for Pop( h) gives:
  $min(0.6175, 0.455, 0.315, 0.4875, 0.95, 0.9)$
  $\times 0.95 = 0.29925 < 0.3$
  $\rightarrow semantic\ incorrectness, validation\ failed$

From the above, it can be concluded that the rule base suffers from semantic incorrectness in the point of view of the student who provided the reference values. However if the difference between 0.61 and 0.7 and also between 0.299 and 0.3 is neglectable, we can conclude that the rule base is near the validation criteria.

## VII. CONCLUSION AND FUTURE WORK

Improving the fidelity and automation of simulations, human behavior modeling is the concern of nowadays research. On this way, in this research a new case study dealing with professor student interaction is defined. The corresponding rule base was constructed by gathering information through specially designed questionnaires. The rule base was mapped to a FPN and through an FPN-based recently presented method was verified to distinguish and refine the structural errors. Afterwards, the semantic errors were distinguished by reasoning the FPN through the dynamic validation.

In the future the presented case-study system will be improved by inserting the professor and student's personal and cultural characteristics within beta-distributions. Also improving the optimality of the certainty factors and truth degrees by artificial intelligence algorithms is not far to imagine if these factors are defined as a fitness function of the student's characteristics such as age, sex, desire for PhD, ranking among classmates. Furthermore the threshold values for transition enabling can be adjusted on the basis of the results of the fuzzy reasoning. Also the concept of truth degrees which dedicate a value to the tokens inside the places may initiate the idea of using colored Petri nets with valued tokens. Also considering priority to fire the enabled transitions in the reachability graph may make us set foot on priority Petri nets.

## REFERENCES

[1] Fei Liu, Ming Yang, Guobing Sun, "Verification of Human Decision Models in Military Simulations", Proceedings of the IEEE 2007, The First Asia International Conference on Modeling & Simulation, pp. 363 - 368

[2] B. H. McNally, "An approach to human behavior modeling in an air force simulation", in Proc. IEEE 2005, Winter Simulation Conference, Orlando, Dec. 2005, pp. 1118-1122.

[3] Naoyuki Kubota, and Kenichiro Nishida, "Prediction of Human Behavior Patterns based on Spiking Neurons", The 15th IEEE International Symposium on Robot and Human Interactive Communication (RO-MAN06), Hatfield, UK, September 6-8, 2006.

[4] Zoran Duric, Wayne D. Gray, Ric Heishman, Fayin Li,Azriel Rosenfeld, Michael J. Schoelles, Christian Schunn, and Harry Wechsler, "Integrating Perceptual and Cognitive Modeling for Adaptive and Intelligent Human–Computer Interaction", Proceedings of the IEEE, Vol. 90, No. 7, pp. 1272 – 1289, July 2002.

[5] Hiroyuki Okuda, Soichiro Hayakawa, Tatsuya Suzuki and Nuio Tsuchida, "Modeling of Human Behavior in Man-Machine Cooperative System Based on Hybrid System Framework", 2007 IEEE International Conference on Robotics and Automation Roma, Italy, 10-14 April 2007.

[6] Wei Ding, Lili Pei, Hongyi Li,Ning Xi, Yuechao Wang, "The Effects of Time Delay of Internet on Characteristics of Human Behaviors", Proceedings of the 2009 IEEE International Conference on Networking, Sensing and Control, Okayama, Japan, March 26-29, 2009.

[7] David W. Dorsey and Michael D. Coovert, "Mathematical Modeling of Decision Making: A Soft and Fuzzy Approach to Capturing Hard Decisions", HUMAN FACTORS, Vol. 45, No. 1, Spring 2003, pp. 117–135.

[8] Michael J. Johnson, Michael McGinnis, "methodology for human decision making using using fuzzy artmap neural networks", Proceedings of the 2002 International Joint Conference on Neural Networks, Vol.3, pp. 2668 – 2673, 2002

[9] Simon C.K. Shiu, James N.K. Liu, Daniel S. Yeung, "An Approach Towards the verification of Fuzzy Hybrid Rule/Frame-based Expert Systems", 12th European Conference on Artificial Intelligence, Published in 1996 by John Wiley & Sons, Ltd

[10] Zhang Hua, Li Rui, Sun Jizhou, "An Emotional Model for Nonverbal Communication based on Fuzzy Dynamic Bayesian Network", Canadian Conference on Electrical and Computer Engineering, pp. 1534 - 1537 2006.

[11] Juanda Lokman Jun-ichi Imai Masahide Kaneko, "Understanding Human Action in Daily Life Scene based on Action Decomposition using Dictionary Terms and Bayesian Network", IEEE 2008 Second International Symposium on Universal Communication.

[12] Wang Li and Wang Mingzhe, "Extraction and Confirmation of Rules for Human Decision Making", IEEE 2009 International Forum on Information Technology and Applications.

[13] Zhuomin Sun, "Multi-Agent Based Modeling: Methods and Techniques for Investigating Human Behaviors", Proceedings of the 2007 IEEE International Conference on Mechatronics and Automation August 5 - 8, 2007, Harbin, China.

[14] Fei Liu, Ming Yang and Peng Shi, "Verification and Validation of Fuzzy Rules-Based Human Behavior Models", 7th International Conference on System Simulation and Scientific Computing, 2008, pp. 813 - 819

[15] M. Suwa, A. Scott, and E. Shortliffe, "An approach to verifying completeness and consistency in a Rule-Based expert system," Technical Report: CS-TR-82-922, pp. 16-21, 1982, Stanford University, Stanford, CA, USA

[16] T. Nguyen, W. Perkings, Y. Laffey, and D. Pecora, "Checking an expert systems knowledge base for consistency and completeness," in Proc. International Joint Conference on Artificial Intelligence, 1985, pp. 375-378.

[17] M. Tousset, "On the consistency of knowledge bases: the COVADIS system", in Proc. European Conference on Artificial Intelligence, 1988, pp. 79-84.

[18] P. Meseguer, "Incremental verification of rule-based expert systems," in Proc. European Conference on Artificial Intelligence, 1992, pp. 840-844.

[19] He, X., Chu, W. C., Yang, H., Yang, S. J.H. "A New Approach to Verify Rule-Based Systems Using Petri Nets". In: IEEE proceeding of 23th Annual International Computer Software and Applications Conference (COMPSAC'99). (1999) 462-467.

[20] T. Murata, "Petri Nets: Properties, Analysis and Application," Proc. IEEE, vol. 77, no. 4, pp. 541-580, 1989.

[21] S. J. H. Yang, J. J.P. Tsai, and C. Chen, "Fuzzy rule base systems verification using high-level Petri nets," IEEE Transactions on Knowledge and Data Engineering, vol. 15, no. 2, pp. 457-473, Mar./Apr. 2003.

[22] A. J. Gonzalez, and M. Murillo, "Validation of human behavioral models," in Proc. Simulation Interoperability Workshop, Mar. 1999.

[23] S. Y. Harmon, and S. M. Youngblood, "Validation of human behavior representations," in Proc. Simulation Interoperability Workshop, Mar. 1999.

[24] Hui-Yu Wang and Shyi-Ming Chen, "Evaluating Students' Answer scripts using Fuzzy Numbers Associated With Degrees of Confidence", IEEE Transactions on Fuzzy Systems, Vol. 16, No. 2, April 2008

[25] Sunghyun Weon And Jinil Kim, "Learning Achievement Evaluation Strategy using Fuzzy Membership Function", 31st ASEE/IEEE Frontiers in Education Conference, October 10 - 13, 2001.

[26] GU Dong-xiao, LIANG Chang-yong, CHEN Wen-en, GU Ya-di, FAN Xin, WU Wei, "Case-based Knowledge Reuse Technology for Questionnaires Design", 4th International Conference on Wireless Communications, Networking and Mobile Computing, 2008, pp. 1 - 4

[27] I. R. Craig and G. L. Burrett, "The Design Of A Human Factors Questionnaire For Cockpit Assessment", An International Conference on

Human Interfaces in Control Rooms, Cockpits and Command Centres, 21 - 23 June 1999

[28] Ayushi Garg and Sumit Singh, "Towards The Adaptive Questionnaire Generation using Soft Computing", World Congress on Nature & Biologically Inspired Computing, 2009, pp. 806 - 811

[29] Shen, V. R. L.: Knowledge Representation using High-Level Fuzzy Petri Nets. IEEE Transactions on Systems, Man, and Cybernetics—Part A: Systems And Humans. Vol. 36, No. 6. (2006) 1220-1227

[30] S.M. Chen, J.S. Ke, and J.F. Chang, "Knowledge Representation using Fuzzy Petri Nets," IEEE Trans. Knowledge and Data Eng., vol. 2, no. 3, pp. 311-319, Sept. 1990.